\documentclass[runningheads]{llncs}
\usepackage{colortbl}
\usepackage{graphicx}
\usepackage{balance}

\usepackage{booktabs} 
\usepackage{tabularx}

\usepackage{xurl}
\usepackage{hyperref}% I'll stop moving; thanks

\usepackage{url}

\usepackage{breakurl}

\usepackage{todonotes} 
\usepackage{subfloat}
\usepackage{xspace}

\usepackage{amssymb}

\usepackage{xcolor}
\definecolor{Gray}{gray}{0.9}

\usepackage{listings}

\usepackage{color} % For custom colors (if not already included)

\definecolor{codegreen}{rgb}{0,0.6,0}
\definecolor{codegray}{rgb}{0.5,0.5,0.5}
\definecolor{codepurple}{rgb}{0.58,0,0.82}
\definecolor{backcolour}{rgb}{0.95,0.95,0.92}

\newcommand{\customsize}{\fontsize{7.5}{10}\selectfont}

\lstdefinestyle{mystyle}{
    backgroundcolor=\color{backcolour},   
    commentstyle=\color{codegreen},
    keywordstyle=\color{magenta},
    numberstyle=\tiny\color{codegray},
    stringstyle=\color{codepurple},
    basicstyle=\ttfamily\customsize,
    breakatwhitespace=false,         
    breaklines=true,                 
    captionpos=b,                    
    keepspaces=true,                 
    numbers=left,                    
    numbersep=5pt,                  
    showspaces=false,                
    showstringspaces=false,
    showtabs=false,                  
    tabsize=2,
    language=SQL
}

\begin{document}
\title{DISL: Fueling Research with A Large Dataset of Solidity Smart Contracts}

%\title{DI\TODO{E}SL: Dataset of Componentized \TODO{Ethereum} Smart Contracts for Large Experiments}
% DISL (DISL Dataset of $514,506$ Unique Smart Contracts)
%\author{First Author\inst{1}\orcidID{0000-1111-2222-3333} \and
%Second Author\inst{2,3}\orcidID{1111-2222-3333-4444} \and
%Third Author\inst{3}\orcidID{2222--3333-4444-5555}}
%
%\authorrunning{Group of Authors}
%\author{Gabriele Morello, Mojtaba Eshghie, Sofia Bobadilla, Martin Monperrus}
\author{Gabriele Morello \and
Mojtaba Eshghie \and
Sofia Bobadilla \and
Martin Monperrus}

\authorrunning{Morello et al.}

\institute{KTH Royal Institute of Technology, Stockholm, Sweden\\
\email{morello@kth.se, eshghie@kth.se, sofbob@kth.se, monperrus@kth.se}}

\maketitle            

\begin{abstract}
The DISL dataset features a collection of $514,506$ unique Solidity files that have been deployed to Ethereum mainnet. It caters to the need for a large and diverse dataset of real-world smart contracts. DISL serves as a resource for developing machine learning systems and for benchmarking software engineering tools designed for smart contracts.
By aggregating every verified smart contract from Etherscan up to January 15, 2024, DISL surpasses existing datasets in size and recency. 
\end{abstract}

\section{Introduction}\label{sec:intro}

% intro to SCs
Smart contracts are self-executing programs running on top of a decentralized ledger~\cite{szabo1996smart}. Their use-cases range from financial applications, governance, gaming to art. 
Solidity stands as one of the most used programming languages for smart contract development.

% security of SC is crucia: 
Security and safety of smart contracts is crucial because their deployment on an immutable ledger makes vulnerabilities and bugs irrevocable, risking significant financial losses. Thus, it is essential to conduct thorough testing, analysis, and repair of buggy smart contract code. Tools and techniques employed for this purpose use both static to dynamic analysis~\cite{SlitherPaper,SGUARD}. 
% low performance, establishing the gap
The performance of these tools however is overshadowed by the number of recent successful attacks performed on real projects deployed on chain~\cite{ArthurGervaisSurveyICSE24}. Recent studies indicate that the reliance on synthetic, small, and non-diverse smart contract datasets fail to represent the complexity of real-world projects. Those small datasets hinder the development and assessment of reliable tools, a core reason for their ineffectiveness~\cite{HowEffectiveAreSCTools,ISSTA2021EmpiricalEvaluation,SlitherPaper}.

% motivation
% why we need a dataset?
A dataset of real-world smart contracts will aid researchers and developers in different tasks, including the production of smart contract development tools. Moreover, as AI-assisted tools gain popularity in smart contract analysis, repair, and synthesis tasks, the demand for large datasets for sake of machine learning tasks on contracts grows increasingly critical~\cite{PEFT,SecurityCodeRecomsSANER2023,GervaisStillManual}. 

Yet, we notice that the current datasets of Solidity smart contracts are limited: they are either outdated (old Solidity versions) and do not reflect real-world deployed contracts (See \autoref{sec:related} for a discussion). 

% solution
To address this problem, we collected DISL, a large dataset of real-world smart contracts. This dataset is useful: 1)~For assessing smart contract analysis tools
2)~For training machine learning and LLM-based tools for smart contracts.

A verified smart contract on Etherscan is a contract whose source code has been uploaded to Etherscan and successfully matched with its compiled code on blockchain to provide transparency and public accessibility to the source.

DISL only includes the verified source code of smart contracts to ensure it comprises solely real contracts in use.
DISL contains the source code for all smart contracts on Etherscan from the genesis block to January 15, 2024. DISL is processed with a deduplication phase in order to remove the Solidity smart contracts that appear thousands of times on Ethereum. 

To sum up our contributions are:
\begin{itemize}
\item The largest dataset of Ethereum smart contracts, containing full source code and metadata satisfying the requirements of a suitable dataset for research and tool development (see \ref{sec:requirements}).
Additionally, as the focus of related datasets (Section \ref{sec:related}) is Solidity contracts, DISL (\emph{raw} collection) with $7,188$ smart contracts in the Vyper programming language, is the largest dataset of Vyper contracts currently available. 
\item A publicly-available version published on Huggingface  accessible for researchers and practitioners: \url{https://huggingface.co/datasets/ASSERT-KTH/DISL}.
% we'll do that later
% \item A long-term archival version available on Zenodo\TODO{Add the url here}. 
\end{itemize}

\section{Dataset Collection Process}\label{sec:collection}

\subsection{Definitions}\label{sec:definitionss}
Here, we clarify the terminology.
\begin{itemize}
    \item \textbf{Deployed contract}: A smart contract address on Ethereum blockchain main network that is associated with a binary code stored in an Ethereum block.

    % we never measured that
    %\item \textbf{Solidity contract}: A single smart contract defined in Solidity language with the \emph{contract} keyword.   
    
    %\item \textbf{Contract source}: A single smart contract defined with  
    \item \textbf{Raw contract}: A concatenated version of all source codes (Solidity files) retrieved from Etherscan for a \emph{deployed contract}. It may contain multiple libraries used within the given contract address. 
    
    %\item \textbf{Decomposed contract}: A \emph{raw contract} that is contract is broken into its constituent components (contracts, libraries, etc). Each component is stored in one separate \emph{Solidity file}. 

    \item \textbf{Solidity file}: The source code files that is used to write Solidity smart contract code. A raw contract may contain several \emph{Solidity files}, which themselves may contain more than one ``contract'' keyword.

\end{itemize}

\subsection{Requirements}\label{sec:requirements}

We want to collect a dataset of smart contracts that meets the following requirements.
\begin{itemize}
    \item The dataset shall contain smart contracts covering a range of different applications (DeFi, art, etc.) 
    \item The dataset shall contain Solidity smart contracts written in recent versions of Solidity (including 2022 onwards). % Therefore, we collect every contract has a verified source code on Etherscan.
    \item The dataset shall be suitable for AI tasks, hence should contain limited duplication. % Hence, inflate the contracts and deuplicate them. Furthermore, collecting all available verified contracts on Etherscan helps with AI tasks that require training or fine-tuning on large datasets.  
    \item The dataset shall be available in a widely-usable format, supported by mainstream dataset platforms.
\end{itemize}

\subsection{Initial dataset.}
%We start our work from Andstor~\cite{andstorHuggingface}.  This dataset contains every deployed Ethereum smart contract until April 1, 2022~\cite{andstorHuggingface}. 
We start our work with the Andstor dataset, which includes every deployed Ethereum smart contract up to April 1, 2022~\cite{andstorHuggingface}.
This dataset contained a total of $2,217,692$ entries of \emph{raw} contracts and $186,397$ lines of deduplicated Solidity files. 
DISL extends Andstor.

\subsection{Collection of most recent contracts.} Next, we use the Google BigQuery database for Ethereum that provides access to blockchain data for analysis~\cite{EthereumBigQuery}. This daily updated database allows exploration of smart contract transactions. We collected smart contracts data starting from the last day of the previously mentioned dataset, April 1, 2022, for all \emph{deployed contracts} that at least have one transaction.

\noindent\begin{minipage}{\textwidth}
\begin{lstlisting}[style=mystyle, caption=Query used for retrieving the dataset, label=lst:sql_query]
 SELECT contracts.address, COUNT(1) AS tx_count
 FROM `bigquery-public-data.crypto_ethereum.contracts` AS contracts
 JOIN `bigquery-public-data.crypto_ethereum.transactions` AS transactions 
 ON (transactions.to_address = contracts.address)
 WHERE transactions.block_timestamp >= TIMESTAMP("2022-04-01")
 GROUP BY contracts.address
 ORDER BY tx_count DESC;
\end{lstlisting}
\end{minipage}

By using Google BigQuery we collected a CSV file with $2,709,030$ entries, where every line has the address of the \emph{deployed contract} and the number of transactions.

Using Etherscan's public APIs, we retrieved the source code for all listed addresses from the data collected on Google BigQuery, successfully acquiring data for $2,660,658$ contracts stored in JSON format . Next, we remove the rows composed of empty JSON files, which indicates when a contract does not have a verified address on Etherscan. This results in what we call a \emph{raw} dataset, consisting of $1,080,579$ rows. 

We merged our parquet files with Andstor dataset to obtain $3,298,271$ rows, each one corresponding to a \emph{deployed contract}.

\subsection{Deduplication}
% Intro paragraph: why do we deduplicate?
The structure of \emph{deployed contracts} is usually a mix of dependencies and project-specific contract definitions that inherit or use the dependencies. The dependencies employed in \emph{deployed contracts} originate from popular smart contract libraries, including OpenZeppelin, Safe, and Provable~\cite{OpenZeppelin,Provable,Safe}. This means source codes in DISL \emph{raw} dataset may contain many times the same Solidity library code. 
Therefore, we perform deduplication (only for \emph{decomposed} collection) to ensure that all dataset entries have unique value.

Using a naive similarity approach would result in discarding many contracts that all use same libraries but still differ in functionality. Thus, per Storhaug et al., we first decompose contracts in separate Solidity files (Section~\ref{sec:definitionss})~\cite{VulnerabilityConstrainedDecoding}. The total number of \emph{Solidity files} after decomposition amounts to $12,931,943$ files.
We filter this collection using the Jaccard similarity index to distinguish the duplicate contracts with a threshold of $90\%$ (the threshold used in ~\cite{adverseEffectsOfDuplicateCode}). After filtering we obtained $514,506$ Solidity files, consolidating our \emph{decomposed} dataset. This means more than $96\%$ in DISL \emph{raw} is duplicate code according to the used similarity scheme.

\begin{table}[!t]
    \centering
    \caption{Details of smart contract metadata in each of the two \emph{raw} and \emph{decomposed} collections of the dataset (each row is a columns in the dataset table)}
    \label{tab:metadata}
    \renewcommand{\arraystretch}{1.5} % Adjusted for better spacing
    \begin{tabular}{lp{3cm}p{6cm}cc}
        \toprule
        & \textbf{Column} & \textbf{Description} & \textbf{Raw} & \textbf{Decomposed} \\
        \midrule
        & Contract name & Contract name & \checkmark & \checkmark \\
        & Contract address & Address of the contract on the Ethereum blockchain & \checkmark & \checkmark \\
        & Language & Language of the contract (Solidity, Vyper) & \checkmark & \checkmark \\
        & Source code & In \emph{raw}, it contains all the code provided by the Etherscan API appended together, in \emph{decomposed} the source code of a single Solidity file & \checkmark & \checkmark \\
        & Compiler version & Version of the compiler used to compile the smart contract & \checkmark & \checkmark \\
        & License type & Name of the license of the smart contract & \checkmark & \checkmark \\
        & ABI & Application Binary Interface of the contract & \checkmark & \(\times\) \\
        & Optimization used & Boolean for running optimization & \checkmark & \(\times\) \\
        & Runs & Number of runs in optimization & \checkmark & \(\times\) \\
        & Constructor arguments & Arguments for the constructor of the smart contract & \checkmark & \(\times\) \\
        & EVM version & Version of the Ethereum Virtual Machine used to compile the contract & \checkmark & \(\times\) \\
        & Library & Name and address of the library used to compile the contract & \checkmark & \(\times\) \\
        & Proxy & Boolean true if the contract is a proxy & \checkmark & \(\times\) \\
        & Implementation & Address of the implementation if the contract is a proxy & \checkmark & \(\times\) \\
        & Swarm source & Address of the source code in Swarm & \checkmark & \(\times\) \\
        & File path & In the \emph{decomposed} dataset, contains the path to the file in the contract structure & \(\times\) & \checkmark \\
        \bottomrule
    \end{tabular}
    % \begin{flushleft}
    %     Note: \checkmark indicates inclusion in the dataset; \(\times\) indicates exclusion.
    % \end{flushleft}
\end{table}

\section{Dataset Content}
As mentioned in Section~\ref{sec:collection}, DISL consists of two collections \emph{raw} and \emph{decomposed}. Table~\ref{tab:overview} provides the overview of DISL. 
It contains more than three millions of deployed smart contracts.
The major outcome of this work is a dataset of $514,506$ real-worl smart contracts (\emph{decomposed} collection), with no duplication (last row of Table~\ref{tab:overview}).
We publish both of these collections as tabular data in Huggingface (\url{https://huggingface.co/datasets/ASSERT-KTH/DISL}).

\begin{table}[t]
    \centering
    \caption{Overview of DISL dataset collections}
    \label{tab:overview}
    \renewcommand{\arraystretch}{1}
    \begin{tabular}{p{5.5cm}c}
        \toprule
        \textbf{} & \textbf{Raw} \\
        \midrule
        \textbf{\# Deployed smart contracts} & 3,298,271 \\
        \textbf{Solidity file count (decomposed)} & 12,931,943  \\
        \textbf{Solidity file count (decomposed, deduplicated)}  & 514,506 \\
        %\textbf{Total Size (Parquet format, raw)} & 20.48 GB \\
        % \textbf{Total Size (Parquet format), decomposed}& 2.32 GB \\
        \bottomrule
    \end{tabular}
    %\begin{flushleft}
    %    Note: Both datasets include comprehensive metadata from Etherscan, enhancing the utility and depth of analysis possible.
    %\end{flushleft}
\end{table}

\textbf{File Content}:  
Each collection is divided into several tabular files to improve manageability. Every file is stored in Parquet format and contains up to $30,000$ rows.

\textbf{Metadata}:
We add metadata provided by Etherscan as columns in the tabular format of the dataset.
Table~\ref{tab:metadata} shows the available metadata.

\textbf{Binary}: The dataset contains all the metadata to actually reconstruct the bytecode representation via compilation. The deployed bytecode of the contract on blockchain is also available with the collected contract addresses. 

\section{Applications}
%The DISL dataset, by virtue of its large collection of verified smart contract source codes, is beneficial in two primary domains of AI-based tool development and the benchmarking of smart contract software engineering tools.

Thanks to its size and nature (only verified smart contracts), the DISL dataset offers significant advantages in two primary areas: AI-based tool development and benchmarking of smart contract software engineering tools.

\emph{Task-specific training of large language models (LLM)} is the process of adapting LLM parameters to improve performance on specific tasks (e.g. code synthesis). Fine-tuning is a proven technique for enhancing the performance of LLMs on code-related tasks~\cite{TooFewBugs,AutomatedStackOverflowSummurization}. DISL is a valuable candidate for such fine-tuning operations as it contains a large smart contract corpora with deduplicated files~\cite{VulnerabilityConstrainedDecoding}

\emph{Benchmarking} involves evaluating the performance of software engineering tools (traditional and AI-based) using a standard dataset to measure their performance. As DISL contains unique, real-world contracts, it is a new valuable benchmarking suite compared to the existing ones. 

\emph{Empirical studies} require real scenarios. However, academic smart contract analysis tools lack  real-life examples~\cite{SecurityCodeRecomsSANER2023,HowEffectiveAreSCTools}. The DISL dataset serves as a valuable resource for studying contracts.

\section{Related Work}\label{sec:related}

\begin{table}[t]
    \centering
    \caption{Related smart contract source datasets and their size (number of source files)}
    \begin{tabular}{p{5.5cm}p{1cm}>{\raggedleft\arraybackslash}p{1.8cm}}
        \toprule
        \textbf{Dataset} & \textbf{Year} & \textbf{Size} \\
        \midrule
         % Andstor (\emph{raw})~\cite{andstorHuggingface} & 2022 & $2,217,692$ \\
         Andstor (\emph{deduplicated})~\cite{andstorHuggingface} & 2022 & $186,397$ \\
         Fiesta (\emph{deduplicated})\footnote{It uses hashes for deduplication.}~\cite{ZellicHuggingface} & 2023 & $149,386$ \\
         Sanctuary~\cite{smart_contract_sanctuary} & 2022 & $144,857$ \\
         SmartBugs-Wild~\cite{SBWild,ThomasEmpricalReview} & 2020 & $47,587$ \\
         Ren et al. ~\cite{ISSTA2021EmpiricalEvaluation,SCBenchmarkSuitesUnified} & 2021 & $46,186$ \\
        DAppSCAN~\cite{DAppSCAN} & 2023 & $39,904$ \\
        \hline
                 DISL (\emph{raw}) & 2024 & $3,298,271$ \\
         DISL (\emph{deduplicated}) & 2024 & $514,506$ \\

        \bottomrule
    \end{tabular}
    \label{tab:datasets}
    \begin{minipage}{8.5cm} % Adjust the width to match your table width
        \scriptsize
        1. It uses hashes for deduplication.
    \end{minipage}
\end{table}

Table~\ref{tab:datasets} summarizes the related work on datasets of smart contracts.

Andstor contains $186,397$ deduplicated Solidity files~\cite{andstorHuggingface}. This dataset served as the initial seed for our work. This dataset uses the same deduplication method as DISL.  %\todo{should we say that our dataset fully includes their raw and inflated(deduplicated for us)?}

The Fiesta dataset contains $149,386$ contracts~\cite{ZellicHuggingface}. It uses hashing to deduplicate contracts. %\todo{This is not correct, it uses hashes to deduplicate, it's Andstor that uses the same method}

Sanctuary has $149,386$ contracts from Ethereum mainnet, among them $71,494$ are unique. According to their website, it has been updated for the last time on the 24th of January 2022. It also contains contracts from testnets of Ethereum and is part of a project that includes other networks. \cite{smart_contract_sanctuary}

SmartBugs-Wild and Meng Ren's dataset, containing $47,587$ and $46,186$ unique smart contracts respectively, have been used in vulnerability analysis and tool evaluation~\cite{SBWild,SCBenchmarkSuitesUnified}. 

DAppSCAN contains 39,904 audited smart contracts. DAppSCAN's relatively small size and focus on audited projects limits its utility for machine learning tasks~\cite{DAppSCAN}.

DISL is the largest dataset of smart contracts to date, with over $3.7$ million \emph{raw} smart contract records and a unique subset of $514,506$ \emph{deduplicated} Solidity files.

\section{Conclusion}
DISL is the largest dataset of smart contract source files at the time of writing. It includes 3,298,271 \emph{raw} and  514,506 \emph{deduplicated} Solidity source files, all taken from real-world, deployed, verified contracts deployed on Ethereum. 
We envision
that DISL will fuel future research on smart contracts.
%the security and reliability of smart contracs.

\bibliographystyle{splncs04}
\bibliography{refs} 

\end{document}